\documentclass[runningheads]{llncs}
\usepackage{amsmath,amssymb}
\usepackage{subcaption}
\usepackage[T1]{fontenc}
\usepackage{graphicx}
\usepackage{caption}

\begin{document}
\title{Continuous Map Matching to Graphs\\ under Travel Time Constraints}
%
%
\author{Yannick Bosch\orcidID{0009-0005-8517-9226
} \and Sabine Storandt\orcidID{0000-0001-5411-3834}}
%
%
\institute{Universität Konstanz
\email{\{yannick.bosch,sabine.storandt\}@uni-konstanz.de}}
\maketitle              
\begin{abstract}
Map matching is a fundamental task in trajectory analysis. The input is an embedded graph with travel time costs, and a trajectory in the form of a spatio-temporal measurement sequence recorded by an entity moving in this graph.
The goal is to associate each measurement with a location in the graph as well as providing a path connecting these locations in order, such that the travel time between consecutive locations obeys the timestamp difference of the respective measurements. This is challenging because measurements are subject to spatial uncertainty and thus it is only known that the entity traversed a disk-shaped area rather than a particular point. Existing methods rely on sampling a finite set of candidate locations within each disk. However, they might fail to identify a feasible map matching path if the true locations are not among the selected candidates. We present the first algorithm for general graphs that considers the entire continuum of possible locations in each disk and hence decides feasibility exactly.  
Under realistic assumptions about the graph and the measurement sequence, we prove that our algorithm runs in   $\mathcal{O}(k n \log n)$, where $k$ is the number of measurements and $n$ the number of nodes in the graph.
We also provide a careful implementation of the algorithm and propose several engineering techniques to improve its practical performance. Experiments on a wide range of trajectories demonstrate the effectiveness of our algorithm.

\keywords{Map matching  \and Shortest path \and Disk arrangement.}
\end{abstract}
\section{Introduction}
Many applications such as traffic monitoring  \cite{goh2012online}, fleet management \cite{efentakis2013towards}, navigation  \cite{white2000some}, and mobility pattern mining  \cite{cho2017basis}  rely on the analysis of large-scale trajectory data collected from GPS and other positioning technologies. However, the measured positions are subject to uncertainty and do not directly reveal the roads on which the movement occurred. For many  downstream tasks, it is therefore  essential to map raw trajectories onto paths in the underlying road network. This gives rise to the following  \textit{map matching problem}:
Given a  graph $G(V,E)$ with a straight-line embedding in $\mathbb{R}^2$ and edge costs $c : E \rightarrow \mathbb{R}^+$ reflecting travel time (interpolated linearly), as well as a spatio-temporal measurement sequence $M$ with $M_i = (p_i, r_i, t_i)$ where $p_i \in \mathbb{R}^2$ denotes the measured position, $r_i \in \mathbb{R}_{\geq 0}$ the positional uncertainty radius, and $t_i \in \mathbb{R}$ the timestamp of the measurement  for $i \in [k]$, the goal is to identify a sequence of locations $l_1, \dots, l_k$  such that
\begin{itemize}
    \item $l_i$ exists in $G$, that is, $\forall i \in [k]: \exists e \in E: l_i \cap e \neq \emptyset$ where $e$ is interpreted as a line segment, 
    \item $l_i$ is contained in the disk $D_i$ of radius $r_i$ centered at $p_i$, that is,  $\forall i \in [k]: ||p_i - l_i||_2 \leq r_i$,  
    \item the travel time from $l_i$ to $l_{i+1}$ in $G(V,E)$ does not exceed the respective timestamp difference, that is, $\forall i \in [k-1]: c(l_i, l_{i+1}) \leq t_{i+1}-t_i$.
\end{itemize}
We call a path in $G$ that traverses $l_1,\dots, l_k$  \textit{feasible} if the travel time between consecutive locations obeys the respective travel time constraint.

\begin{figure}
    \centering
    \includegraphics[width=0.66\linewidth]{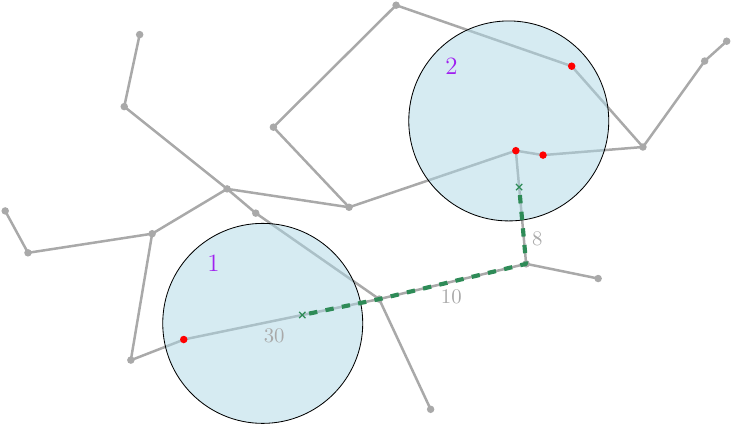}
    \caption{True movement trajectory (dashed green) in a road network (gray) with recorded measurement disks (blue). The timestamps of measurements 1 and 2 are 20 seconds apart. However, if only network nodes are considered as location candidates (red), there is no path from disk 1 to disk 2 that obeys this travel time constraint. The provided travel time edge costs indicate that the shortest path would indeed take 48 seconds. Thus, the feasible solution is missed.}
    \label{fig:mm}
\end{figure}

Existing methods  \cite{lou2009map,eisner2011algorithms,hu2016if} sample a finite set of candidate locations from each disk $D_i$ and use those as nodes in a layered directed acyclic graph (DAG), with one layer per disk. Edges connect nodes in consecutive layers whenever the shortest-path cost in $G$ between the corresponding locations satisfies the travel time constraint induced by their timestamp difference. A path from the first to the last layer then yields a valid map matching solution. However, sampling-based methods are not guaranteed to detect a feasible solution, since the true locations may not be among the sampled candidates, see Figure \ref{fig:mm} for an illustration.
If high sample rates (along the edges contained in $D_i$) are used to overcome this issue, the DAG size and the number of necessary shortest path computations get very large and thus the running time increases significantly.

In this paper, we propose the first exact algorithm for deciding whether a feasible map matching solution exists. We consider the continuum of locations in $D_i \cap G$ as candidates for $l_i$, rather than restricting them to a finite set of samples. Despite the infinite number of candidates, we prove that our algorithm runs in polynomial time and can be engineered to perform very  well in practice.

\subsection{Related Work}
There exist a multitude of approaches for map matching, see \cite{huang2021survey,jiang2022driving} for recent surveys. This includes several machine learning based methods \cite{zhao2019deepmm} as well as geometric algorithms \cite{gudmundsson2024map,chambers2020map}. However, none of these methods is guaranteed to identify a feasible matching path under travel time constraints, and additionally they either suffer from large training effort or are rather slow in practice.

The DAG algorithm described in the introduction is often equipped with more sophisticated  location candidate selection and routing strategies to improve the solution quality in practice \cite{lou2009map,custers2022physically,hu2016if,bast2018sparse}. But those are not sufficient to ensure that a feasible path obeying  travel time constraints is detected.

For the special case of path graphs, an exact algorithm (called COMMA) for map matching under travel time constraints was proposed in \cite{bosch25}. COMMA allows to continuously place locations inside the disks without any discretizations. It was shown that DAG based methods fail to reliably compute a feasible mapping path even in this restricted scenario. In contrast, COMMA always computed the correct solution and was also  significantly faster. Generalizing COMMA to be applicable to arbitrary graphs was left as an open problem.
\subsection{Contribution}
In this paper, we present the first exact algorithm for map matching under travel time constraints in general graphs. We prove that whenever a feasible solution exists, there is a canonical feasible path that can be represented as a concatenation of homotopic shortest paths in the arrangement induced by the positional measurement uncertainty disks. This characterization allows us to compute a feasible map matching in polynomial time, more precisely in  $\mathcal{O}(k^4 n^4 \log (kn))$, where $k$ is the number of measurements and $n$ the number of nodes in the graph.
We also show that under realistic assumptions about the graph and the measurement sequence, the running time reduces to  $\mathcal{O}(k n \log n)$.  Furthermore, we discuss engineering techniques to  accelerate our algorithm in practice. We conduct an extensive experimental evaluation on synthetic and real-world trajectories. The results validate our methodology and also show that our new algorithm clearly outperforms the DAG baseline with respect to running time.

\section{Continuous Map Matching  Algorithm}
In this section, we describe our novel map matching algorithm. While DAG based approaches restrict mapping locations to a finite set of samples in each disk, our algorithm considers the continuum of locations inside each disk. Clearly, the DAG based methodology, which creates a node for each location candidate, can not deal with this scenario. We provide several insights about the structure of feasible paths and use those to efficiently search for a valid solution. The main idea is to classify paths by their intersection structure with the boundaries of the measurement disks.

\subsection{Boundary intersections and homotopic paths}
Let $B_i$ be the boundary circle of  disk $D_i$ of radius $r_i$ centered at $p_i$. We define $\phi_i := B_i \cap E$ to be the set of intersection points of $B_i$ with edges in $G(V,E)$. To avoid discussion of degenerated cases, we assume that the intersection points do not coincide with endpoints of edges, and that there is no point in $G$ which is a boundary intersection point for multiple disks, that is $\phi_i \cap \phi_j = \emptyset$ for all $i,j \in [k], i \neq j$.  We use $\Phi = \bigcup_{i=1}^k \phi_i$ to refer to the set of all boundary circle intersection points. Given this set, we subdivide the edges $e \in E$ at each of their contained boundary points $e \cap \Phi$ and call the resulting node set $V^+ = V \cup \Phi$ and the new edge set $E^+$. We refer to the graph with node set $V^+$ and edge set $E^+$ as $G^+$. Based on this augmentation, the cells in the disk arrangement subdivide $G^+$ into edge-disjoint subgraphs.

\begin{figure}
    \centering
    \includegraphics[width=0.6\linewidth]{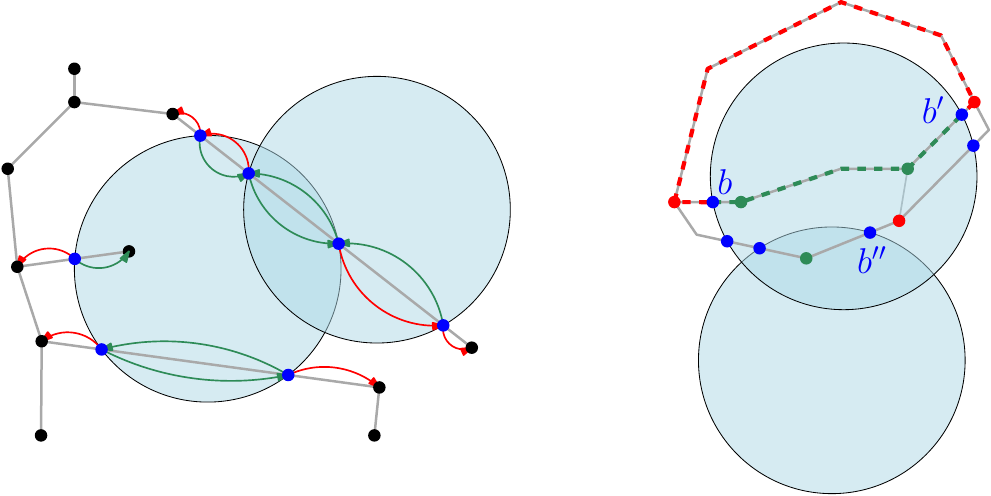}
    \caption{Left: Intersection points (blue) of graph edges (gray) and disk boundaries (black). Red arrows point from intersection point $b$ to $out(b)$ and green arrows from $b$ to $in(b)$. Right: The homotopic in-in-path from $b$ to $b'$ is marked by the dashed green line, the homotopic out-out-path by the dashed red line. Although the path via $b''$ might be shorter, it includes  additional intersection points, which is not allowed. A homotopic in-out or out-in path does not exist from $b$ to $b'$. However, there is a homotopic  out-in path from $b''$ to $b$ and from $b''$ to $b'$.}
    \label{fig:homotopic}
\end{figure}

Our goal is to show that a feasible map matching path can be represented by a sequence of so-called homotopic paths between boundary points.
\begin{definition}[Homotopic path]
Given two locations $s,t$ that exist in  $G^+$ (i.e., they are points on some edge segment), we call a path $\pi(s,t)$ from $s$ to $t$ in $G$ a \emph{homotopic path} if its interior contains no point from $\Phi$. We use $\pi^{\gamma:\iota}(s,t)$ to denote paths from $s$ to $t$ that contain no point from $\bigcup_{i=\gamma}^\iota \phi_i$.
\end{definition}
Thus, a homotopic path does not cross any boundary of the cell arrangement induced by the disks, and a homotopic $s$-$t$ path exists only if $s$ and $t$ lie in the same cell. For $\pi^{\gamma:\iota}(s,t)$, these properties are defined with respect to the arrangement induced by the disks $D_\gamma,\dots,D_\iota$. For a point $b\in\phi_i$, let $e=\{v,b\}$ and $e'=\{b,w\}\in E^+$ be its two incident edges, with $v\notin D_i$ and $w\in D_i$. We refer to $v$ as $out(b)$ and to $w$ as $in(b)$. A path between two boundary points $b,b'\in\Phi$ is called \emph{in-out-homotopic} if its first edge is $\{b,in(b)\}$, its last edge is $\{out(b'),b'\}$, and the path from $b$ to $b'$ is homotopic. The other path types are defined analogously. See Figure \ref{fig:homotopic} for illustrations.
 We will now show that the search for a feasible map matching path can be restricted to concatenations of homotopic shortest paths in the disk arrangement.
\begin{theorem}\label{thm:concat}
    If there exists a feasible location sequence $l_1, l_2, \dots, l_k$   for given $G^+$ and $M$, there exists a matching path which is a concatenation of shortest homotopic paths between points in $\Phi$.
\end{theorem}
\begin{proof}
Let $\psi$ be a feasible matching path, and let $B_\psi=b_1,\dots,b_q$ denote the intersection points of $\psi$ with disk boundaries, in the order in which they occur along $\psi$. First, assume that all locations are mapped after $b_1$ and before $b_q$. Let $\psi'$ be a path from $b_1$ to $b_q$ in $G^+$ obtained by concatenating shortest homotopic paths between consecutive boundary points. The path type is determined by the role of the respective intersection points. If $b_j$ is the $m$th intersection with disk $D_i$ along $\psi$ and $m$ is odd, we have an in-x-path; otherwise, we have an out-x-path. For $b_{j+1}$, the corresponding types are x-out-path and x-in-path, respectively. We claim that $\psi'$ is also a feasible matching path. Let $l_i$ be the location to which measurement $M_i$ is mapped on $\psi$, and let $b_j,b_{j+1}$ be the boundary points between which $l_i$ lies. Since there are no points from $\Phi$ on the subpath of $\psi$ from $b_j$ to $b_{j+1}$, this subpath cannot be shorter than the corresponding homotopic path in $\psi'$. Let $t_\psi(b_j,l_i)$ denote the travel time from $b_j$ to $l_i$ along $\psi$. We choose $l_i'$ on $\psi'$ such that $t_{\psi'}(b_j,l_i')=t_\psi(b_j,l_i)$. If this location lies beyond $b_{j+1}$, we instead set $l_i'=b_{j+1}$. Thus, $t_{\psi'}(b_j,l_i')\leq t_\psi(b_j,l_i)$ and $t_{\psi'}(l_i',b_{j+1})\leq t_\psi(l_i,b_{j+1})$. Consequently, $t_{\psi'}(l_i',l_{i+1}')\leq t_\psi(l_i,l_{i+1})$, so all travel-time constraints are satisfied. Moreover, the construction preserves the order of the locations along the path, and since $l_i'$ lies on a homotopic path, we have $l_i'\in D_i$. Hence, $l_1',\dots,l_q'$ is a feasible mapping sequence and $\psi'$ is a feasible matching path.
If $\psi$ maps locations before $b_1$ or after $b_q$, these locations lie on homotopic paths in the respective cells. We can therefore map the corresponding prefix or suffix directly onto $b_1$ or $b_q$, respectively, without violating the order or time constraints.
Finally, if $\psi$ contains no boundary point, all disks overlap and $\psi$ is fully contained in the cell defined by their intersection. In this case, we can simply choose any boundary point of this cell and map all locations onto it.
\end{proof}
The characterization of feasible mapping paths as concatenations of shortest homotopic paths between boundary points suggests the following algorithm: First, compute shortest homotopic paths between all pairs of boundary points. Then, enumerate all possible concatenations and test each for feasibility using the map matching algorithm for path graphs from \cite{bosch25}. However, this naive approach has a running time exponential in $|\Phi|$. We therefore seek a stronger characterization that significantly reduces the number of candidate paths.
\subsection{A polynomial time algorithm}
\begin{figure}
    \centering
    \includegraphics[width=0.6\linewidth]{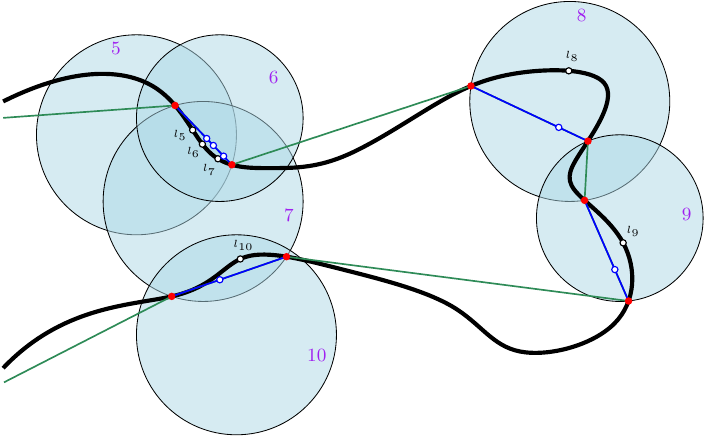}
    \caption{Illustration of a feasible mapping path (black) with location placement. The red marked boundary points enclose the location encoding intervals. The canonical path representation is shown in blue and green, where blue segments indicate homotopic shortest paths and green segments general shortest paths.}
    \label{fig:canonical}
\end{figure}

To make the search more efficient, we observe that every feasible mapping path $\psi$ contains at most $k$ homotopic subpaths on which some location $l_i$, $i\in[k]$, is mapped. Let $in_i$ and $out_i$ denote the boundary points enclosing the subpath associated with $l_i$. If $out_i\neq in_{i+1}$, we can connect them by a shortest path in $G^+$ without homotopy constraints. The resulting path remains feasible, since shortening the path between $l_i$ and $l_{i+1}$ cannot violate their time constraint. Combined with Theorem \ref{thm:concat}, this yields a canonical representation of a feasible path that preserves feasibility as follows.
\begin{definition}[Canonical path representation]
Given a feasible mapping path $\psi$. Let $I=\{(in_1,out_1), (in_2, out_2), \dots\}$ denote the set of intervals induced by intersections of $\psi$ with disk boundaries that contain mapped locations. Then the canonical representation of $\psi$ alternates between homotopic shortest paths between $in_i$ and $out_i$ and shortest paths between $out_i$ and $in_{i+1}$.
\end{definition}
Figure \ref{fig:canonical} illustrates this concept. The canonical representation allows us to reduce the  path candidates to be checked to combinations of at most $k$ homotopic subpaths, using shortest path interpolations between them where necessary. However, as now all concatenations of two homotopic subpaths are possible, this would demand to check all permutations. 

To avoid this, we also consider the location sets covered by each homotopic path. Along a feasible mapping path, each location-containing interval $I_j$ contains a contiguous subsequence of locations $l_\gamma,\dots,l_\iota$, and the next interval $I_{j+1}$ must contain $l_{\iota+1}$, the first location not covered by $I_j$. Thus, for a given instance, we aim to compute all feasible intervals together with all possible location subsets.
For this purpose, we introduce the concept of a homotopic location group. A sequence of consecutive locations $l_\gamma,\dots,l_\iota$ forms a homotopy group $g_{\gamma:\iota}$ if the disks $D_\gamma,\dots,D_\iota$ have a non-empty intersection cell, which can therefore contain all these locations simultaneously. For a homotopy group $g_{\gamma:\iota}$, let $C := \bigcap_{j=\gamma}^{\iota} D_j$ be the corresponding cell and $C\cap\Phi$ the set of boundary intersection points contained in it. For every pair $s,t\in C\cap\Phi$, we compute $\pi^{\gamma:\iota}(s,t)$, the shortest homotopic path between $s$ and $t$ that remains within $C$. If $s$ or $t$ lies on the boundary of $C$, the path must traverse the node $in(s)$ or $in(t)$, respectively. We collect all such paths in a path set $\Pi$, storing for each its endpoints, travel-time cost, and associated location set $l_\gamma,\dots,l_\iota$.

Given the full path set $\Pi$, our goal is to identify a feasible matching path or certify that none exists. We partition $\Pi$ according to $l_\iota$, the last location covered by each path, and denote the resulting partitions by $\Pi_\iota$, $\iota\in[k]$. For each $\Pi_\iota$, let $Out_\iota$ be the union of the path endpoints. We then sweep through the partitions in order and assign each $out\in Out_\iota$ the largest available time budget $\beta_{\iota+1}(out)$ for reaching $l_{\iota+1}$. Once $l_\iota$ is determined, this budget is $\beta_{\iota+1}(out)=t(l_{\iota+1})-t(l_\iota)-c(l_\iota,out)$, where the first term is the timestamp difference and the second is the travel time from $l_\iota$ to $out$. For $\iota=1$, we can map $l_1$ directly onto $out$, yielding $\beta_2(out)=t(l_2)-t(l_1)$.
For $\iota>1$, consider a path $\pi\in\Pi_\iota$ with location set $l_\gamma,\dots,l_\iota$. If $\gamma=1$, we again map all locations onto $out$ and set $\beta_{\iota+1}(out)=t(l_{\iota+1})-t(l_\iota)$ (or $0$ if $\iota=k$). Otherwise, every possible predecessor $\pi'$ must lie in $\Pi_{\gamma-1}$. For each $\pi'\in\Pi_{\gamma-1}$, we compute the shortest path in $G^+$ from its endpoint $out'$ to the start $in$ of $\pi$, with cost $c$. The remaining budget for reaching $l_\gamma$ on $\pi$ is then $\beta_\gamma(out')-c$. If this is negative, the concatenation is infeasible. Otherwise, we place $l_\gamma$ on $\pi$ greedily so that this budget is fully spent, or on $out$ if $out$ is reachable, and greedily place the remaining locations as close to $out$ as possible while respecting the time constraints. This determines the travel time from $l_\iota$ to $out$ and hence $\beta_{\iota+1}(out)$.
For each $out\in Out_\iota$, we retain only the path in $\Pi_\iota$ that maximizes its non-negative budget, together with its location placement and predecessor $\pi'$, and discard all other paths. If any partition becomes empty, we immediately conclude that no feasible matching path exists. Conversely, if some path in $\Pi_k$ admits a valid placement of all its associated locations, a feasible matching path exists, and we can recover it by backtracking through the stored predecessor paths and location placements.
\begin{theorem}\label{them:correctness}
    The sweep algorithm correctly decides if there exists a feasible mapping path.
\end{theorem}
\begin{proof}
    Let $\psi$ denote the  canonical representation of a feasible matching path. Let $I$ be the location-containing intervals of $\psi$  and let $in_i, out_i \in \Phi$ denote the entry and exit points of interval $I_i \in I$. Let $l(out_i)$ be the last location placed on $\psi$ before $out_i$. We will prove inductively that our algorithm identifies for each $i \in [|I|]$ a feasible matching path $\psi_i$ for the locations $l_1, \dots, l(out_i)$ that ends in $out_i$ and has a travel time that does not exceed the travel time to reach $out_i$ on $\psi$. This implies, that we could always concatenate  $\psi_i$ with the suffix of $\psi$ that starts at $out_i$ to obtain a feasible matching path. Thus, we will identify a feasible matching path for all measurements in the end.

    For the first interval $in_1,out_1$, there needs to be a homotopy group that contains $l_1, \dots, l(out_1)$ and $out_1$ needs to be a boundary point within the intersection cell of the arrangement of the respective disks. As our algorithm enumerates all homotopic paths for each homotopy group, the path $out_1$ to $out_1$ will be among them with a travel time of $0$. Thus, when checking $\Pi_{l(out_1)}$, all locations up to $l(out_1)$ are assigned to the single node path $out_1$. This path will not be pruned as with a travel time of $0$ it does not use up any budget to reach the next location in the sequence. Thus our hypothesis is true for the base case.

    Now, for the inductive step, we assume our hypothesis is true up to $out_i$ and now consider $out_{i+1}$. Based on our construction of $\Pi$, we know that  $\Pi_{l(out_{i+1})}$, needs to contain a homotopic shortest path from $in_{i+1}$ to $out_{i+1}$ with location set $l_\gamma, \dots, l_\iota$ where $l_\gamma$ is the successor of $l(out_i)$ and $l_\iota = l(out_{i+1})$. When processing this path, we thus consider $\Pi_{l(out_i)}$ and especially $out_i$ as potential predecessor. By the induction hypothesis, $out_i$ is assigned with a budget to reach $l_\gamma$ which is at least as large as the time travel difference of $l(out_i)$ and $l_\gamma$ on $\psi$ and which encodes a feasible matching path up to location $l(out_i)$ which ends in $out_i$. This implies, that we can concatenate this path with the homotopic shortest path from $in_{i+1}$ to $out_{i+1}$ contained in  $\Pi_{l(out_{i+1})}$. The only reason why this path might not be maintained by our algorithm is that there is yet another feasible path identified that also covers the locations up to $l(out_{i+1})$ but reaches $out_{i+1}$ even earlier. Thus, we conclude that our induction hypothesis is valid and therefore our algorithm correctly computes a feasible matching path if there exists one.
\end{proof}
\begin{theorem}\label{thm:runningtime}
 The sweep algorithm runs in $\mathcal{O}(k^4 n^4 \log(kn))$, where $n$ is the number of nodes in $G(V,E)$ and $k$ is the number of measurements. 
\end{theorem}
\begin{proof}
   he algorithm conducts the following three steps:
     \begin{enumerate}
         \item Computing $\Phi$, the set of boundary intersection points, and constructing $G^+$
         \item Computing the homotopy groups and $\Pi$, the set of homotopic shortest paths
         \item Conducting the sweep
     \end{enumerate}
      For computing $\Phi$, we naively check for all disks with which edges they intersect in time $\mathcal{O}(mk)$. To construct $G^+$, multiple intersection points on one edge need to be sorted first. As each edge can have a most $2k$ such intersection points, this step takes $\mathcal{O}(m k \log k)$. The resulting graph $G^+$ has $|V^+| = |V| + |\Phi| \leq n + 2km$ nodes and $|E^+| \leq (2k+1)m$ edges. 
      
      For homotopy group computation, we observe that there are at most $k^2$ such groups, as each  of them is defined by the  intersection cell of a contiguous subsequence of disks. The boundary of a face/cell in any  disk arrangement consists of $\mathcal{O}(k)$ arcs \cite{aronov2011complexity}. Consequently, this is also the time to check whether a homotopic group can be extended by the next disk in the sequence.  To compute $\Pi$, we need to find for each homotopy group $g_{\gamma:\iota}$ the set of points $S \subseteq\Phi$ that are contained in the intersection cell of the respective disk arrangement. This can be achieved by starting with $\Phi$ and, for each subsequent disk, deleting the points that are not contained in the respective intersection cell anymore. Thus, the overall time to compute the homotopy groups and the point sets $S$  is in $\mathcal{O}(k^3 \Phi)$. Afterwards, we compute for each group $g_{\gamma:\iota}$ the shortest paths between each pair of points in $S$ under the restriction that no point in $\Phi' = \bigcup_{i=\gamma}^\iota \phi_i$ occurs in its interior. This can be achieved by running Dijkstra from each node $s \in S$ but blocking the nodes in $\Phi'$ (except as start or endpoint). Blocking takes $\mathcal{O}(|\Phi|)$ per group and each Dijkstra run takes  $T_{dijk} \in \mathcal{O}(|V^+| \log |V^+| + |E^+|)$. Thus, in total, the running time for computing the path set is in $\mathcal{O}(k^3 |\Phi| + k^2 |\Phi| T_{dijk}$). The resulting path set size is bounded by $|\Pi| \in \mathcal{O}(k^2 |\Phi|^2)$. 
      
      For the sweep, we observe that  $|\Pi_\iota|  \in \mathcal{O}(\iota |\Phi|^2)$ as there are at most $\iota$ homotopic groups with largest location index $\iota$. Processing a path $\pi \in \Pi_\iota$ requires to look at a path set $\Pi_{\iota' < \iota}$, which after processing has size at most $|\Phi|$ (at most one path for each point in $Out_{\iota'}$). Computing the largest remaining budget at $in$, the start node of $\pi$, requires one run of Dijkstra's algorithm simultaneously from all $out' \in  Out_{\iota'}$ and deducting  the cost from the respective budgets $\beta_\iota(out')$. Note that we only need to run this Dijkstra once for all requests to $\Pi_{\iota'}$ if we store the respective results, that is, the maximum remaining budget at any graph node. After looking up the budget for the current $in$ node, the locations in the location set associated with  $\pi$ can be placed greedily on $\pi$  in time proportional to the size of this set.  However, within the algorithm, we only care about the  placement of the last location on this path. Thus, we can simply use the timestamp difference of the first and the last location in the set to directly place the last location on $\pi$. Thus, processing $\pi$ requires constant time. Only after termination of the algorithm the remaining intermediate locations are placed on the respective subpaths of the final mapping path,  which incurs a cost of $\mathcal{O}(k)$ once. Thus, processing $\Pi_\iota$ takes time $\mathcal{O}(\iota |\Phi|^2)$ for processing each contained path and $\mathcal{O}(T_{dijk})$ for the Dijkstra run from all surviving nodes in $Out_\iota$. Thus the whole sweep  runs in $\mathcal{O}(k^2 |\Phi|^2 + k T_{dijk})$.

      Under worst-case assumptions, we have $|\Phi| \in \mathcal{O}(kn^2)$ as well as  $|V^+|, |E^+| \in \mathcal{O}(kn^2)$ and consequently $T_{dijk} \in \mathcal{O}(kn^2 \log{kn})$. This results in an overall running time of $\mathcal{O}(k^4 n^ 4 \log(kn))$.
\end{proof}

\subsection{Analysis under Realistic Assumptions} \label{sec:realistic}
Although our novel map matching algorithm has polynomial running time, its practical applicability is questionable due to the high polynomial degree. However, the worst-case analysis relies on several unrealistic assumptions: the input graph may have quadratically many edges, every disk boundary may intersect every edge, all possible homotopy groups may exist, and each may contribute the maximum possible number of homotopic paths. In this section, we provide a more realistic analysis. To simplify discussion, we also assume $k\ll n$.
\smallskip\\\textbf{Assumption 1: Sparse movement networks.}
Movement networks such as road and pedestrian networks are typically sparse, so we expect $m\in\mathcal{O}(n)$ in practice. This gives $|V^+|,|E^+|\in\mathcal{O}(kn)$, $T_{dijk}\in\mathcal{O}(kn\log n)$, and $|\Phi|\in\mathcal{O}(kn)$. Thus, the   time for computing $\Pi$ and   the sweep reduces to $\mathcal{O}(k^4n^2\log n)$.
\smallskip\\\textbf{Assumption 2: Few boundary intersection points.}
For trajectories obtained, e.g., from GPS measurements, uncertainty radii are typically at most a few hundred meters, so each disk covers only a small portion of the network. It is therefore reasonable to assume a constant number of boundary intersections per disk, yielding $|\Phi|\in\mathcal{O}(k)$. Together with Assumption 1, this gives $|V^+|,|E^+|\in\mathcal{O}(n)$ and $T_{dijk}\in\mathcal{O}(n\log n)$. Substituting these bounds into the running times from the proof of Theorem \ref{thm:runningtime} yields   $\mathcal{O}(k^3n\log n)$.
\smallskip\\\textbf{Assumption 3: Few overlapping disks.}
Measurements are often taken at fixed time intervals, e.g., every 30 seconds. Many overlapping disks thus require either large uncertainty, slow movement, or frequent returns to similar locations. Assuming none of these applies, we expect each disk to overlap with only $\mathcal{O}(1)$ other disks. This reduces the number of possible homotopy groups from $\mathcal{O}(k^2)$ to $\mathcal{O}(k)$, which can then be computed in linear time. Combined with Assumptions 1 and 2, the number of boundary points in each homotopy group's intersection cell is also $\mathcal{O}(1)$ and the overall running time reduces to $\mathcal{O}(k^2+kT_{dijk})\in\mathcal{O}(kn\log n)$.

We investigate the validity of these assumptions and the resulting running-time bounds in the experimental evaluation.

\section{Implementation and Engineering}
In this section, we describe  engineering techniques that can substantially improve practical performance without changing the theoretical running time.
\smallskip\\\textbf{Boundary intersection point computation.}
We need to compute  intersections between disk boundaries and graph edges. Rather than testing every disk-edge pair, we index the  edges in a spatial data structure such as an AABB-tree and, for each disk, test only edges intersecting its smallest enclosing square.
\smallskip\\\textbf{Homotopy blocks.}
For a homotopy group $g_{\gamma:\iota}$, the relevant boundary-point set $S$ may contain points from disks $D_\tau$ with $\tau\ll\gamma$ or $\tau\gg\iota$. However, if an intermediate disk is completely disjoint from all preceding measurement disks, such as $D_8$ in Figure \ref{fig:canonical}, boundary points from earlier disks do not impact the location placements and thus can be ignored. We therefore partition the measurements into blocks, ending a block whenever the next disk does not overlap any previous disk. Homotopic shortest paths are then constructed using only disks from the same block, potentially reducing their number substantially.
\smallskip\\\textbf{Bounded Dijkstra runs.}
During the sweep, we run Dijkstra simultaneously from all nodes in $Out_\iota$ by initializing each node with its associated time budget. Edge costs are subtracted from these budgets while propagating them, retaining the maximum budget at each node. Since nodes with negative budgets cannot belong to a feasible matching path, we terminate the search once no active node has a positive budget. This can substantially limit the explored part of the graph.

\section{Experimental Results}
The above algorithm was implemented in C++, and all experiments were conducted using a single core of an AMD Ryzen Threadripper 3970X processor.

\begin{figure}[]
    \centering
    \begin{subfigure}[b]{0.3\textwidth}
        \centering
        \includegraphics[width=\linewidth]{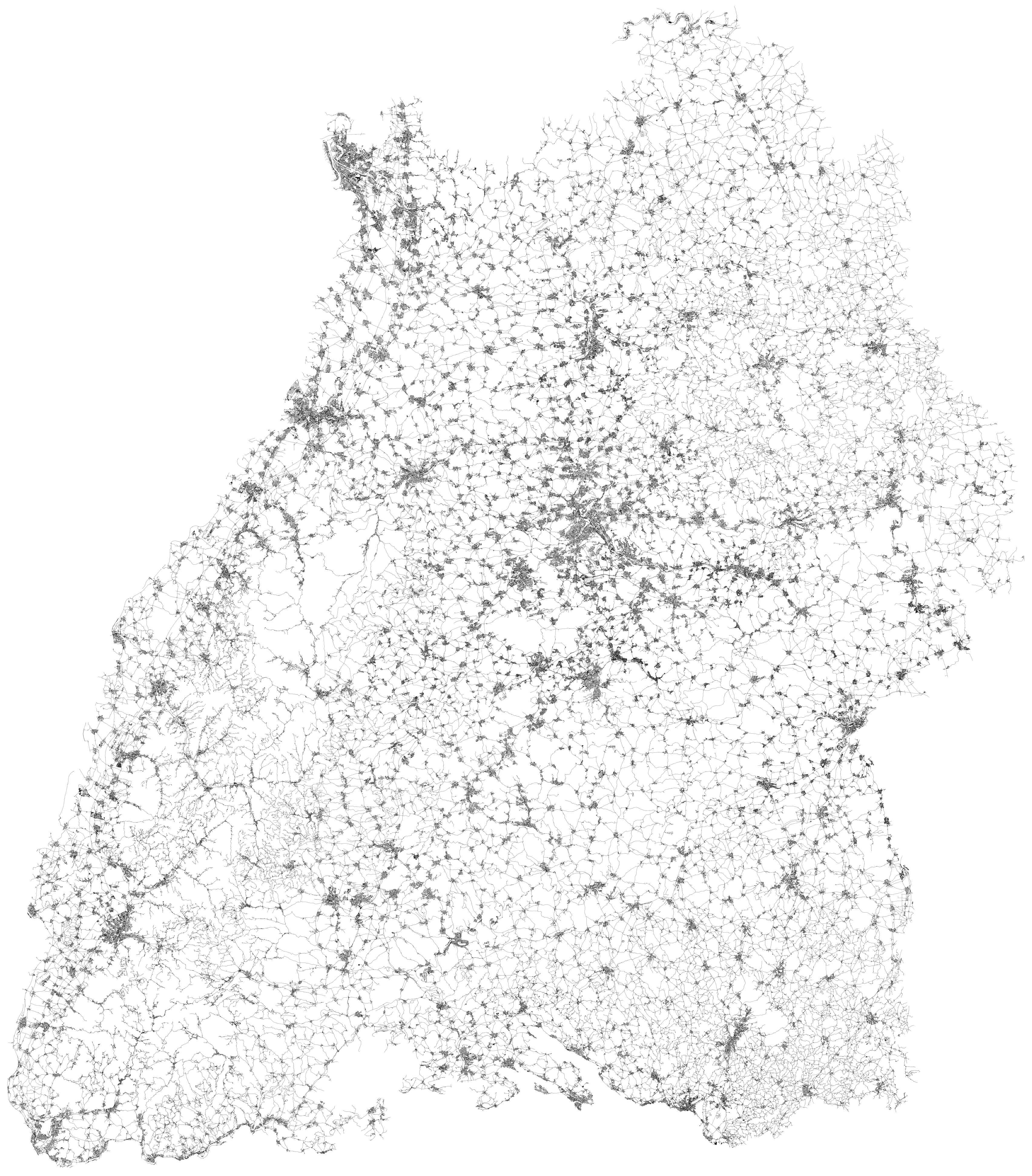}
    \end{subfigure}
    \quad
    \begin{subfigure}[b]{0.45\textwidth}
        \centering
        \includegraphics[width=\linewidth]{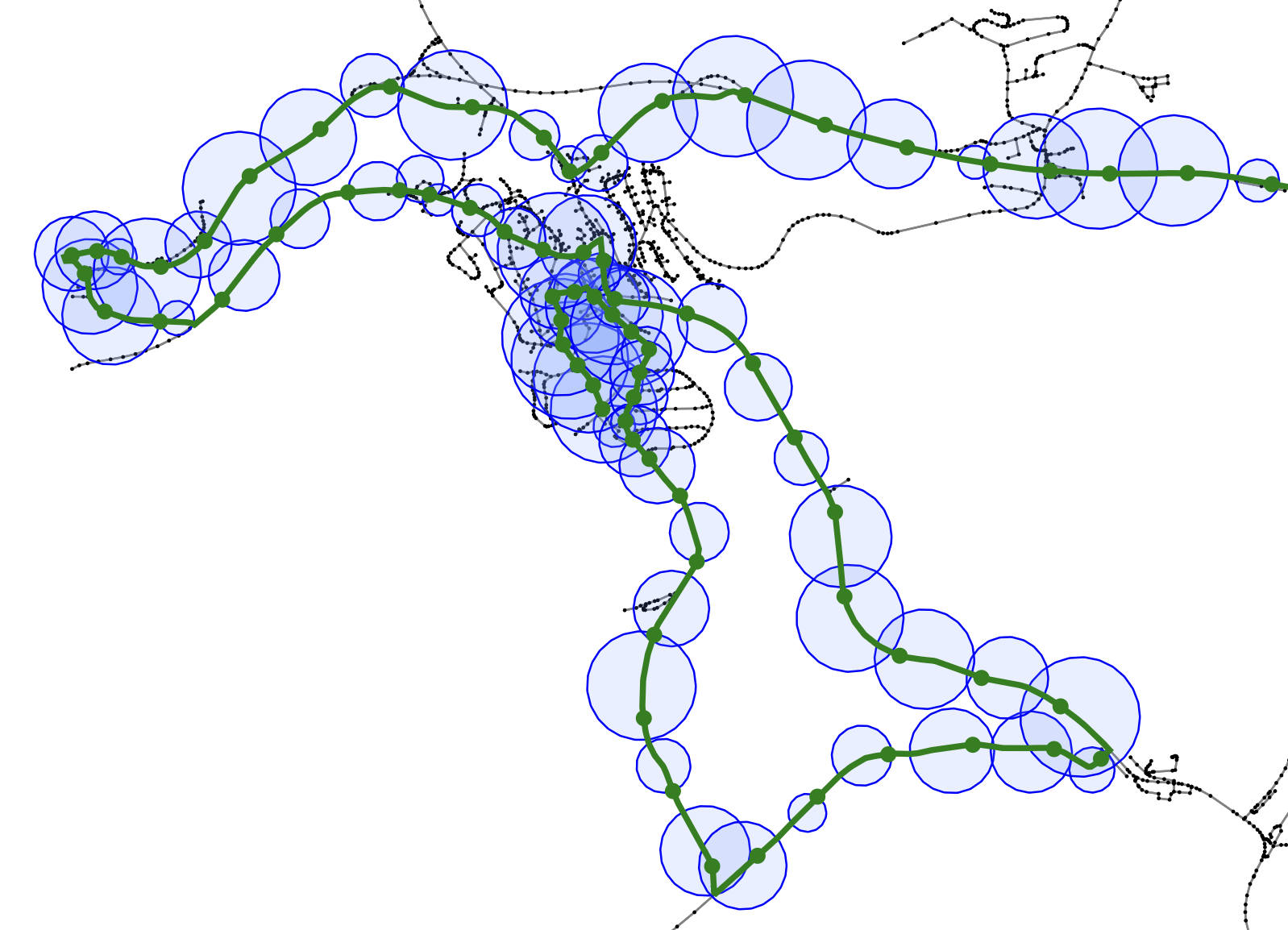}
    \end{subfigure}
    \caption{Road network of Baden-Württemberg alongside a solved example instance stemming from a DFS. The whole network contains approximately $5$ million nodes and $10$ million edges.}
    \label{fig:bw}
\end{figure}

For the evaluation, we extracted instances from the road network of Baden-Württemberg (see Figure \ref{fig:bw}). The instances were generated either from shortest paths in the network or by means of a depth-first traversal. Measurement points were then sampled along the resulting paths according to a specified travel-time rate. For each measurement point, an uncertainty  radius between 50 and 200 meters was selected uniformly at random. The measurement points were then perturbed within their respective disks. This construction guarantees that the original points constitute a feasible solution for every generated instance. If not otherwise specified we assume the described values as the default.

\subsubsection{Running Time Performance}
\begin{figure}[h]
    \centering
    \includegraphics[width=0.49\linewidth]{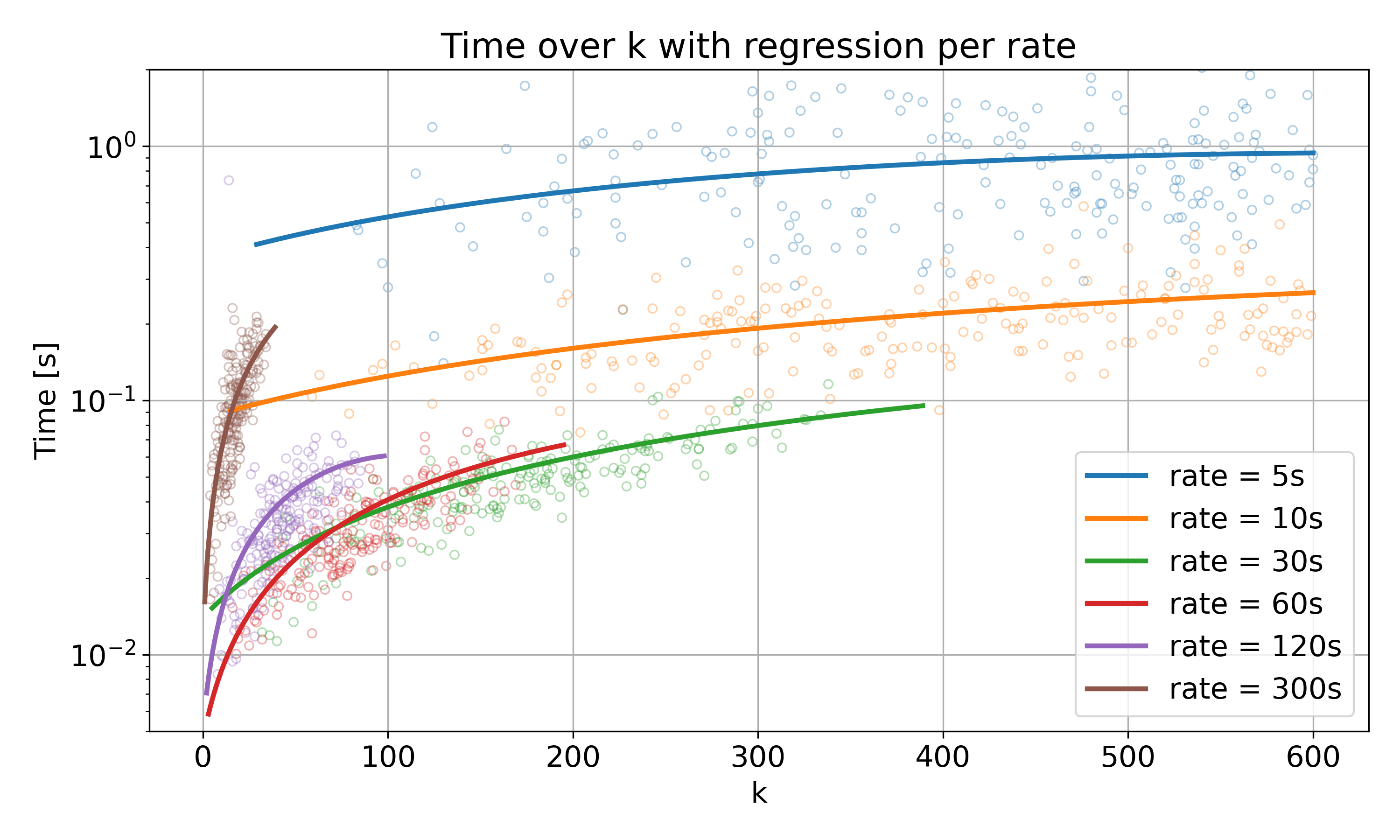}  
    \hfill
    \includegraphics[width=0.49\linewidth]{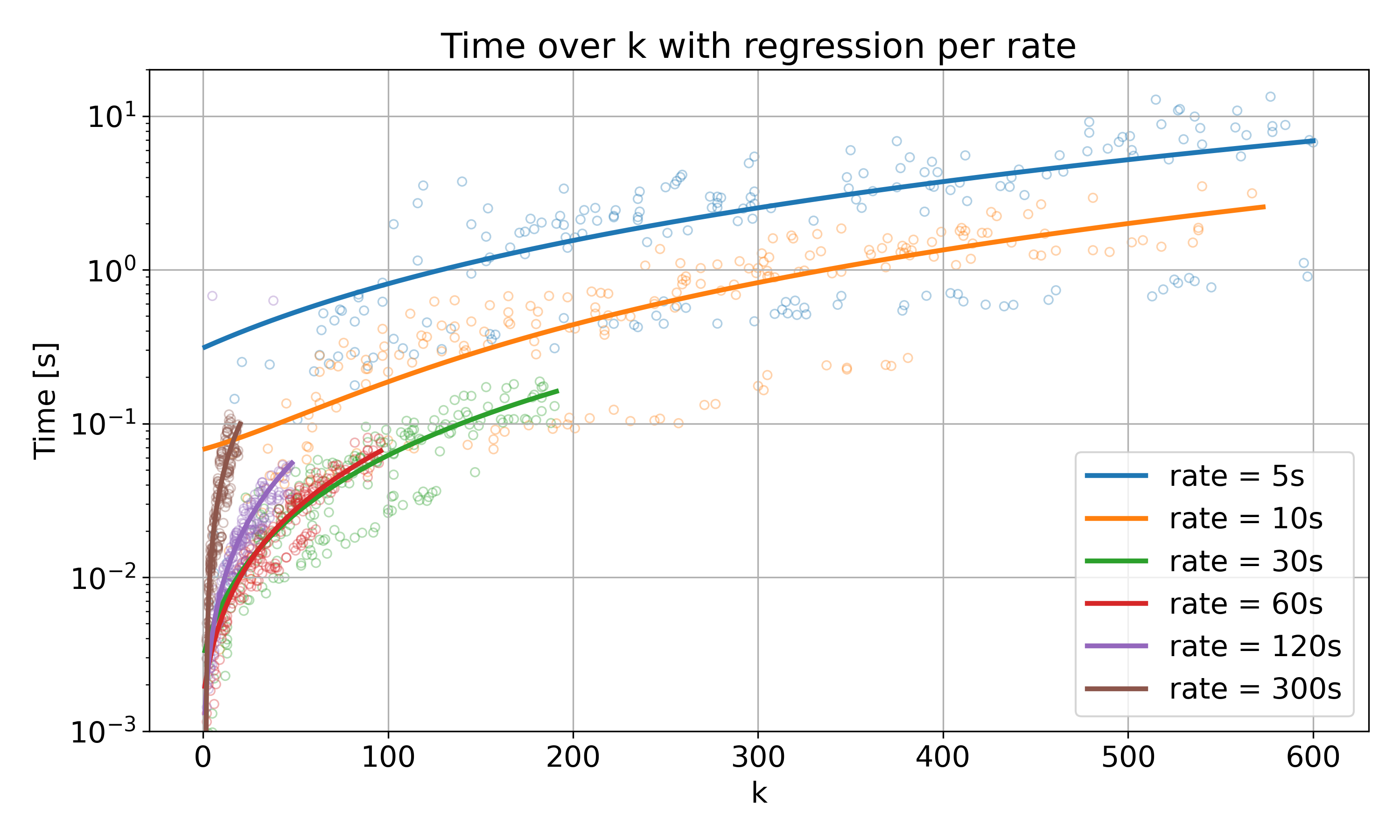}  
    \caption{Algorithm running time for instances extracted from shortest paths on the left and DFS on the right. For each sampling rate around 10000 instances were evaluated, the points shown only represent a subset thereof for visual clarity.}
    \label{fig:RTSpDfs}
\end{figure}

As shown in Figure \ref{fig:RTSpDfs}, our algorithm achieves low running times, averaging below one second for shortest-path instances and below ten seconds for DFS instances.

Table \ref{tab:PRT} highlights the particular difficulty of the DFS instances, which exhibit an average maximum overlap of approximately $38$. It also provides insight into the running-time distribution across the three main phases of our algorithm. When the overlap is large and the time interval is small, computing the homotopy groups accounts for the majority of the running time. In contrast, for larger  time intervals, most of the computation is spent on the sweep, which is expected as the Dijkstra runs get more expensive.
\begin{table}[h]
    \centering
    \caption{Average running time percentages and average maximum overlap for different time  rates on the   DFS paths data from Figure \ref{fig:RTSpDfs}.}
    \label{tab:PRT}
    \begin{tabular}{|r||r|r|r||r|}
    \hline
        Rate [s] & Construct $G^+$ [\%] & H. Groups [\%] & Sweep [\%] & Overlap \\
        \hline
        5   &0.523812	&60.127062&	39.349126&	37.613010 \\
        10  &2.444198	&62.971002&	34.584800&	18.437664\\
        60	&16.273135	&32.095759&	51.631106&	2.780176\\
        300 &4.658992	&2.503395	&92.837613& 1.002527\\
    \hline
    \end{tabular}
\end{table}

\subsubsection{Comparison to DAG}
In this experiment, we implemented the DAG-based improvement described in the introduction, which samples edges within the disks. 
\begin{figure}[h]
    \centering
    \includegraphics[width=0.6\linewidth]{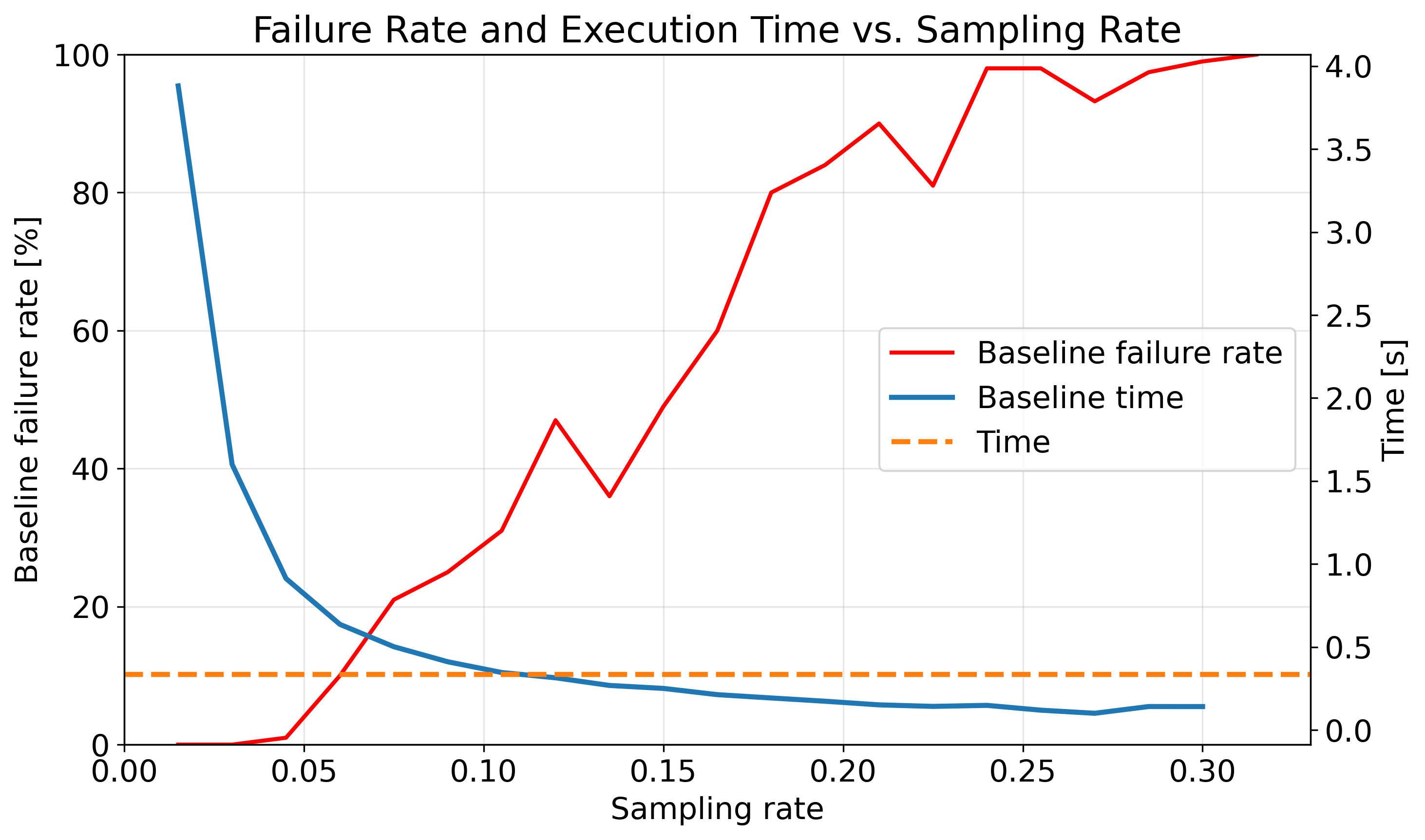}  
    \caption{Running time and failure rate of the DAG baseline over the sampling rate. The plot is averaged over $100$ instances of size $k=200$ of the metropolitan region of Stuttgart, with uncertainty radii ranging from $50$ to $75$ meters, travel time rate of $5$ seconds. The sampling rate ranges from $0.015$ to $0.315$ seconds. For the former the average number of sampling locations per disk is $2076$.}
    \label{fig:BaselineFR}
\end{figure}
However, as shown in Figure \ref{fig:BaselineFR}, this approach has limited effectiveness: even for smaller instances as $k=200$, a very low sampling rate is not guaranteed to find a valid solution. In contrast, our approach has constant running time, as it is independent of the sampling rate, and consistently produces valid solutions, as all considered instances were feasible.

\newpage

\subsubsection*{Real World Evaluation}
\begin{figure}
    \centering
    \includegraphics[width=0.6\linewidth]{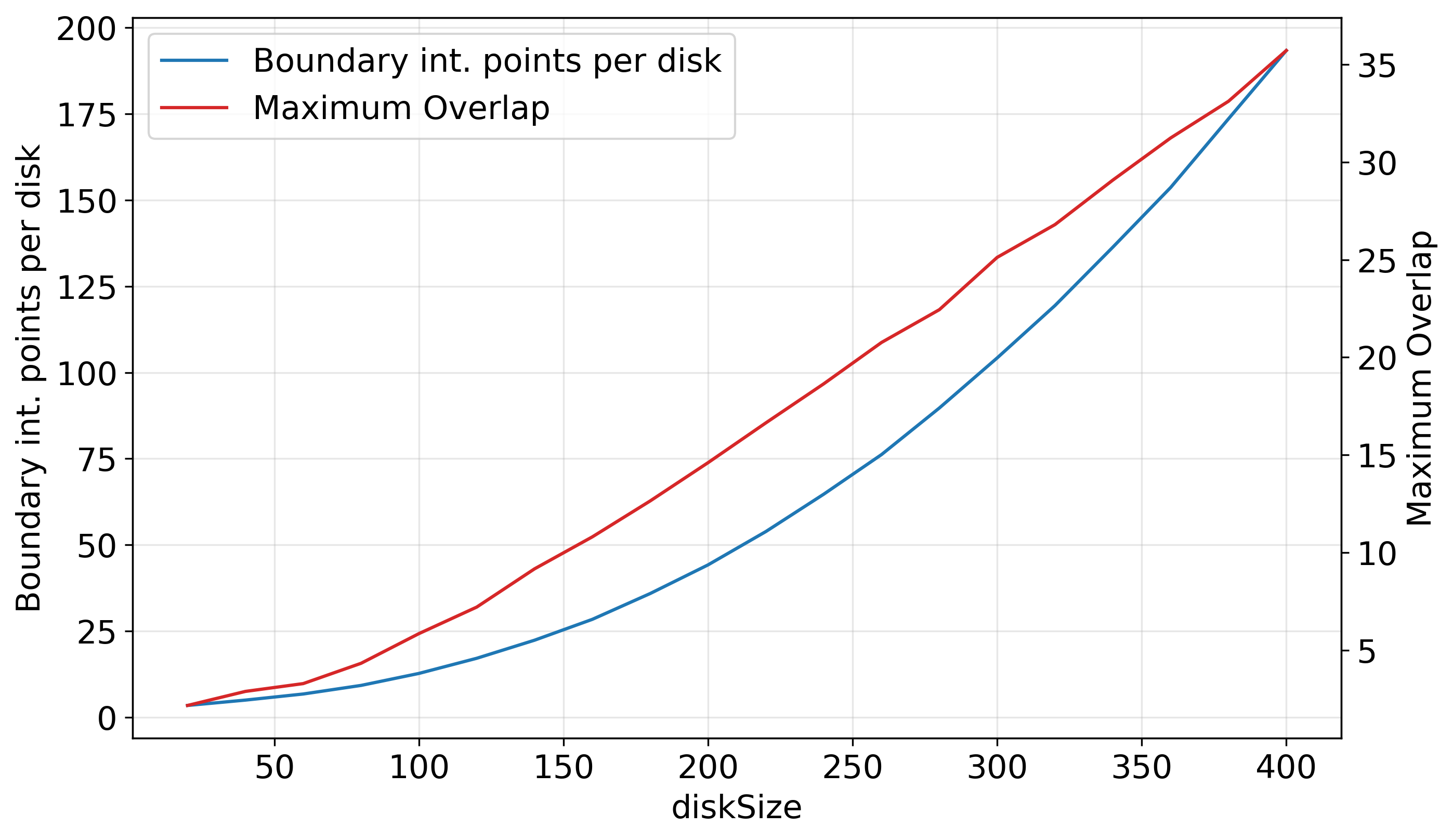}  
    \caption{Average boundary intersection points of our algorithm and average maximum overlap plotted over increasing disk size. 25 instances of length $k=200$ are considered with a travel time rate of 30s between measurements. }
    \label{fig:DE}
\end{figure}

Figure \ref{fig:DE} further demonstrates that realistic instances as described in Section \ref{sec:realistic}, with a measurement rate of $30$s and a disk radius of $100$ meters, exhibit relatively few boundary intersection points per disk and low maximum overlap, confirming our assumptions. This is also reflected in the running times reported in Figure \ref{fig:RTSpDfs}.
\vspace{-3em}
\subsubsection*{Evaluation on Real Trajectories}
\begin{figure}[h]
    \centering
    \includegraphics[width=0.6\linewidth]{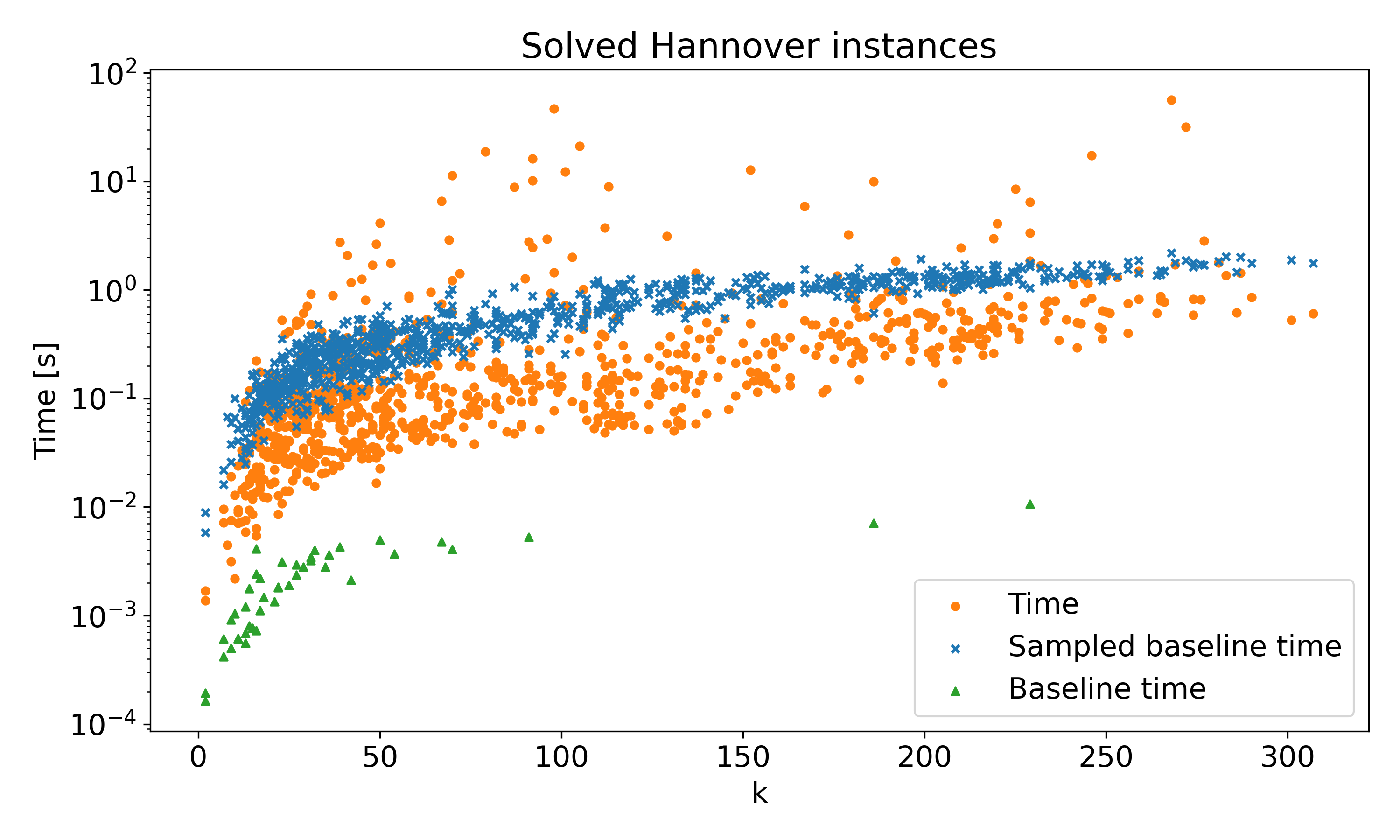}  
    \caption{Running times of solved instances of the GPS trajectory dataset of the Leibnitz Universität Hannover \cite{hannoverGPS}. The instances were processed to contain measurements every 5 seconds. The measurement uncertainty radii are set at 75 meters. The sampling-based DAG baseline used a sampling rate of $0.025$ seconds.}
    \label{fig:Hannover}
\end{figure}

Finally, Figure \ref{fig:Hannover} demonstrates that our algorithm can efficiently solve real-world instances. 

The processed dataset stems from car trips around the city of Hannover \cite{hannoverGPS} and contains 1204 instances, of which only $3.72\%$ were solvable using the naive DAG baseline and $76.02\%$ using the sampling-based DAG baseline with on average $1951$ sampling locations per disk. In contrast, our approach produced solutions for $76.48\%$ of the instances. The others were certified to be infeasible. Enlarging the measurement uncertainty radii increases the percentage of feasible instances further. 

Furthermore, we analyzed the properties of said instances regarding the analysis of Section \ref{sec:realistic}. We found that even with a measurement rate of $5$s and a disk radius of $75$ meters, the instances exhibit relatively few boundary intersection points per disk ($28.2$) and a low maximum overlap ($17.59$), confirming our assumptions. This is also reflected in the average running time of these instances of $0.63$ seconds for an average number of about $105$ measurements per instance.

\section{Conclusions and Future Work}
We have shown that deciding whether a measurement sequence can be mapped onto a path in an underlying movement network while respecting travel time constraints is possible in polynomial time, and that feasible paths can be obtained efficiently in practice. Future work could investigate whether additional preprocessing can further improve query times, for example through advanced circle-segment intersection data structures \cite{afshani2025circle} or shortest path speed-up techniques \cite{eisner2011algorithms}. Another direction is to extend our sweep algorithm to more sophisticated travel-time models that account, for example, for acceleration \cite{custers2022physically}.

%
%
\bibliographystyle{splncs04}
\bibliography{references.bib}

\end{document}